\documentclass[12pt]{iopart}
\usepackage{graphicx}
\usepackage{dcolumn}
\usepackage{bm}
\usepackage{epsf}
\usepackage{amssymb}
\usepackage{subfigure}
\usepackage{cancel}
\DeclareGraphicsExtensions{.eps}
\begin{document}

%
%
%
%

\title[High Capacity Quantum Key Distribution]{High Capacity Quantum Key Distribution via Hyperentangled Degrees of Freedom }

\author{David S. Simon$^{1,2}$ and Alexander V. Sergienko$^{2,3,4}$}

\address{1. Dept. of Physics and Astronomy, Stonehill College, 320 Washington Street, Easton, MA 02357, USA}

\address{2. Department of Electrical and Computer Engineering, Boston
University, 8 Saint Mary's St., Boston, MA 02215, USA}

\address{3. Photonics Center, Boston
University, 8 Saint Mary's St., Boston, MA 02215, USA}

\address{4. Dept. of Physics, Boston University, 590 Commonwealth
Ave., Boston, MA 02215, USA}

\eads{simond@bu.edu, alexserg@bu.edu}

\begin{abstract}
Quantum key distribution (QKD) has long been a promising area for application of quantum effects toward solving real-world problems. But two major obstacles have
stood in the way of widespread applications:  low secure key generation rates and short achievable operating distances. In this paper, a new physical mechanism
for dealing with the first of these problems is proposed: interplay between different degrees of freedom in a hyperentangled system (parametric down conversion)
is used to increase the Hilbert space dimension available for key generation while maintaining security. Polarization-based Bell tests provide security checking,
while orbital angular momentum (OAM) and total angular momentum (TAM) provide higher key generation rate. Whether to measure TAM or OAM is decided randomly on
each trial. The concurrent non-commutativity of TAM with OAM and polarization provides the physical basis for quantum security. TAM measurements link
polarization to OAM, so that if the legitimate participants measure OAM while the eavesdropper measures TAM (or vice-versa), then polarization entanglement is
lost, revealing the eavesdropper. In contrast to other OAM-based QKD methods, complex active switching between OAM bases is not required; instead, passive
switching by beam splitters combined with much simpler active switching between polarization bases makes implementation at high OAM more practical.
\end{abstract}


\submitto{\NJP}
\maketitle

\section{Introduction}

\subsection{Quantum key distribution} In quantum key distribution (QKD) two experimenters (Alice and Bob) generate a shared cryptographic key, using quantum mechanics
to guarantee that an eavesdropper (Eve) cannot obtain significant information about the key without being revealed. Commonly, for optical QKD schemes, key bits
are derived from photon polarization. This can be done by having Alice prepare for Bob a single photon in a randomly chosen polarization state known only to her
(BB84 protocol \cite{bb84}), or by Alice and Bob each receiving from a common source half of an entangled photon pair (Ekert protocol \cite{e91}). Either way,
polarization measurements by Eve produce detectable disturbances. Each photon is prepared in one of two non-orthogonal, mutually unbiased bases. Eve, intercepting a
photon traveling to Bob, must guess which basis to measure in; if she measures in the same basis as the two legitimate participants, she acquires full
information without detection. However, half the time she guesses the wrong basis, ensuring that her outcome is uncorrelated with Alice's; she then obtains no
useful information and simultaneously exposes herself to detection by randomizing Bob's results. This occurs because the polarization operators in the two
non-orthogonal bases are not mutually commuting. Exchanging results for a subset of measurements, Alice and Bob see the decrease in correlation between their
polarizations, revealing Eve's actions. For fiber systems, phase is often used instead of polarization, but principle remains the same.

There have been two principal obstacles to widespread application of QKD outside of research labs. First, most methods have been limited in the distances over
which they can operate; the simplest single-photon or weak coherent state approaches, for example, are generally limited to tens of kilometers before photon
losses introduce unacceptable levels of error.

Second, most approaches to QKD with optical systems have used polarization or phase as the variables from which cryptographic key segments are generated.
However, polarization can normally only encode one qubit per photon, unless substantial extra complications to the apparatus are added to allow for qutrit or ququart exploitation. Similarly, it is difficult at a practical level to
increase the number of dimensions of the states encoded by phase beyond two, or at most, four. It would therefore be desirable to find a more practical means of encoding
high dimensional states into a photon. This would increase the rate of key generation by allowing more than one bit of cryptographic key to be shared between the legitimate users of the system per
exchanged photon.

As a means of increasing the key rate, there has been much interest (see \cite{grob,gruneisen,malik,simonfib} and references therein) in using the photon's
orbital angular momentum (OAM) instead of polarization. The range of applications of states with OAM, such as Laguerre-Gauss states, to both classical and
quantum communication has been rapidly expanding; see for example \cite{fickler2,krenn}. OAM is quantized, $L_z=l_z\hbar$, with integer topological charge $l_z$.
There is no fundamental upper limit to the value of $l_z$, so the alphabet size or effective Hilbert space dimension, $N$, is in principle unbounded. Using a
range of $l_z$ values from $-l_0$ to $+l_0$, each photon can generate up to $\log_2N=\log_2(2l_0+1)$ bits of cryptographic key. OAM was first successfully used
\cite{grob} to generate a quantum key by means of the three-dimensional qutrit space spanned by $l_z=0,\pm 1$. However, switching between unbiased, non-commuting
bases in a higher-dimensional OAM space is now needed. This basis switching is much more difficult for OAM than for polarization and the difficulty grows
with increasing basis size. Therefore the apparatus complexity and experimental difficulty increase rapidly with growing $N$.

In this paper, we wish to propose a means for increasing the number of secret key bits generated per transmitted photon, while avoiding the increasingly
difficult basis modulations required by other schemes when going to higher Hilbert space dimensions. In the next section, we introduce a new physical mechanism
for key generation that will allow a simpler experimental route to this goal. The approach we propose makes use of OAM for increased Hilbert space dimension, but
here the OAM is employed in a fundamentally different manner than in all previous methods. In particular, we will make use of its joint entanglement with
polarization, arranging the setup in such a way that the polarization is able to serve as a signal of attempts at eavesdropping on the OAM. Conceptually, rather
than switching between two nonorthogonal bases in the space of \emph{orbital} angular momentum, we switch between two nonorthogonal bases in the larger space of
{\it total} angular momentum. This is much easier to accomplish because only the measurement basis of the photon polarization needs to be actively modulated.

\subsection{Quantum Nondemolition Measurements}

The idea of using OAM to generate a secret key while only doing security-enforcing basis modulations in polarization seems to have an immediate problem. The two
variables commute, so that one may be measured without disturbing the other. For example, the OAM eigenvalue, $l_z$, may be obtained by performing an ideal
quantum nondemolition (QND) measurement \cite{brag1,brag2,thorne,unruh}. This in principal causes no disturbance to the polarization or spin.

In practice, the situation is a little more complicated. Practical execution of QND requires nonlinear optical processes such as Kerr nonlinearity; but it has
been shown that the physics of nonlinear interaction ensures that QND measurement of OAM will cause some disturbance to the signal's polarization state
\cite{lin}. This both reveals Eve's presence and destroys the information she was attempting to obtain, since it prevents Alice and Bob from agreeing reliably on
a shifted key. Furthermore, the low amplitudes of nonlinear processes guarantee low efficiency at the single photon level; only a small fraction of the photons
will participate in the interaction, allowing Eve to determine only a small fraction of the OAM values.

But these considerations are specific to the case of QND measurements via Kerr nonlinearities. It may be possible that Eve has an advanced technology that allows
her to make QND measurements of $l_z$ by some other, as yet unknown, means. There is no fundamental principal, to our knowledge, that guarantees that other such
QND technologies \emph{must} cause a similar disturbance to polarization when applied to OAM. Thus, we wish to avoid this problem by arranging a fundamental
linkage between the polarization and OAM that will cause QND measurements of one variable to disturb the other, \emph{independent of the physical mechanism used
to make the measurement}. We propose a means of accomplishing this in the following sections.

\subsection{Hyperentangled QKD} We propose a high-dimensional OAM-based QKD scheme that requires no random switching between OAM bases. For full security, it is necessary to treat both variables, polarization and OAM, in a
fully quantum manner. The goal is to do this in such a way that active basis modulations are only needed in polarization, not in OAM. This is achieved by adding
a third variable, the \emph{total} angular momentum (TAM) $\hat{\bm J}$ about the propagation axis, and then allowing a random choice between measuring $\hat{\bm
J_z}=\hat{\bm L_z}+\hat{\bm S_z}$ or measuring $\hat{\bm L_z}$. $\hat{\bm J}$ provides a linkage between the spin $\hat{\bm S}$ (which determines the circular
polarization state) and the OAM $\hat{\bm L}$, that allows the desired goal to be achieved.

The principal idea is to separate key generation and security into \emph{different} degrees of freedom; however these variables must be closely linked in such a way that unauthorized
measurements of one variable produce detectable signatures in the other. We will use OAM and TAM in tandem for key generation (due to their high dimensionality and
subsequently high key-generation capacity), while employing polarization for security checks (due to the ease of alternating between polarization bases). This is
possible because spontaneous parametric down conversion (SPDC) supplies photon pairs {\it hyperentangled} \cite{kwiat,atature,barreiro,dossantos,nagali,karimi1}
in polarization and OAM (among other variables). Hyperentanglement in SPDC has found applications in recent years ranging from quantum interferometry
\cite{atature} and quantum imaging \cite{bonato} to quantum cryptography and dense coding \cite{barreiro2,gauthier}. Polarization-OAM hyperentangled
states have also been used for ultra-sensitive angular measurements \cite{ambrosio}.

Previous uses of OAM in conjunction with polarization for QKD applications \cite{barreiro2,gauthier} make use of the two variables in a sort of
parallel, non-overlapping manner: the measurement of one variable has no effect on the other. The components of the two variables are simply appended to each
other to form a vector with more components, thereby expanding the relevant state space to higher dimension. A complicated procedure of basis switching must still
be carried out in this higher-dimensional space in order to ensure security. In contrast, in the current paper hyperentanglement is used in a more intrinsic manner;
the pair of entangled variables are partially overlapping in the sense that they can be \emph{either} independent \emph{or} perfectly correlated with each other,
depending on whether or not a third variable has also been measured. In this way, measurements by the eavesdropper on one variable become apparent through loss
of entanglement in a second variable due to the pairwise noncommutativity of the first two variables with the third. The enforcement of security measures is then
greatly simplified at high dimensions, since passive switching between the two measurement variables is technically much simpler than active switching between
large basis sets for a single variable.

In addition to the technical simplification of QKD at high dimensions, the proposed procedure is interesting for several other reasons.  Use of the
noncommutativity of $\bm {\hat J}$ with $\bm {\hat L}$ and $\bm {\hat S}$ for quantum communication applications seems to be largely unexplored, as is the use in
QKD of pairs of \emph{different vector operators} in place of pairs of \emph{different components of the same vector operator}. Further, the use of angular
momentum erasure (see section (\ref{proceduresection})) to maintain polarization entanglement has not previously been proposed and may be interesting in its own
right as the basis for angular-momentum-based analogs of quantum eraser and delayed choice experiments.

The proposed approach relies on the fact that although OAM and polarization commute with each other, neither of them commutes with the TAM. To be explicit, the
commutators of $\bm {\hat L}$ and $\bm {\hat S}$ among themselves are given by:
\begin{eqnarray} \left[ \hat L_i,\hat L_j\right] &=&
i\sum_k\epsilon_{ijk}\hbar L_k\\  \left[ \hat S_i,\hat S_j\right] &=&
i\sum_k\epsilon_{ijk}\hbar S_k \\
\left[ \hat L_i,\hat S_j\right] &=& 0
, \end{eqnarray} so that:
\begin{eqnarray} \left[ \hat L_i,\hat J_j\right] &=& \; \left[ \hat L_i,\hat L_j+S_j\right] \; =\; \left[ \hat L_i,\hat L_j\right] +\cancel{\left[ \hat L_i,
\hat S_j\right]} \\ &=& i\sum_k\epsilon_{ijk}\hbar L_k\\
\left[ \hat S_i,\hat J_j\right] &=& \; \left[ \hat S_i,\hat L_j+S_j\right] \; =\;  \cancel{\left[ \hat S_i,\hat L_j\right]} +\left[ \hat S_i,\hat S_j\right]
\\ &=& i\sum_k\epsilon_{ijk}\hbar S_k ,
.\end{eqnarray} This lack of commutativity means that TAM measurements provide an indirect linkage between $\bm {\hat L}$ and $\bm {\hat S}$, ; we make use of
this linkage in the following.

The specific procedure to be proposed in the next section allows Alice and Bob a random choice between measuring the eigenvalues of either the OAM ($\hat{\bm
L_z}$) or the TAM, ($\hat{\bm J_z}$). Only trials on which Alice and Bob both measure the same variable are kept; on these, the photon spin (polarization) state
remains entangled. If Eve measures a variable different than Alice and Bob did, the spin wavefunction collapses into a definite polarization state, which will be
detectable by a Bell-type test on polarization. This occurs because once the values of $\hat{\bm L_z}$ and $\hat{\bm J_z}$ have both been measured, the value of
$\hat{\bm S_z}=\hat{\bm J_z}-\hat{\bm L_z}$ can be determined as well. Subsequent measurements of the photon's linear polarization in the $x-y$ plane will then
be affected by this sequence of measurements.

%

The higher dimension of the OAM state space
increases the number of key bits generated per photon; this is done without \emph{any} additional
active modulation beyond what is needed in the usual polarization-based protocols. There is no
upper limit to the number of bits possible in principle, although there are of course practical limits.
$N$ can potentially be scaled up to very large size with little
additional effort as long as sources with high values of entangled angular momenta and OAM sorters
that work over a large enough range are both available.  The range of achievable $l$
values for entangled photons has rapidly grown in recent years \cite{romero,fickler}. Measures
must be taken to guarantee that the range being used for the alphabet
has a flat spectrum; otherwise Eve can use the
differing probabilities to gain information about the key. This equalization, however, can be easily achieved,
for example by using extra
OAM sorters followed by filters with different transmission rates. The span of values
that can be sorted by a single sorter has also grown, though more slowly \cite{leach2,karimi,guo,berk,sluss,lavery,osullivan2}.

\begin{figure}
\centering
\includegraphics[totalheight=2.6in]{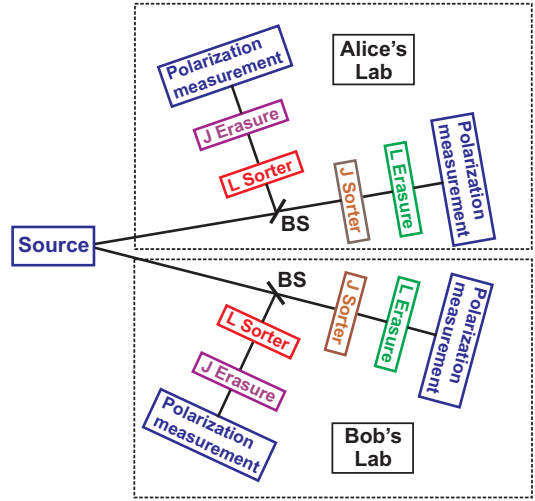}
\caption{\textit{Each participant randomly measures either $\bm {\hat L_z}$ or $\bm {\hat J_z}$ via nondestructive
sorting. After sorting is done in
one of these variables, the information about the other variable is erased (see figure \ref{erasure}). Regardless of which variable is sorted and which is erased,
the polarization is undisturbed and available for measurement.}}
\label{sortfig}
\end{figure}

\section{Setup and Procedure}\label{proceduresection}

\subsection{The Setup} 
\begin{figure}
\centering
\includegraphics[totalheight=2.4in]{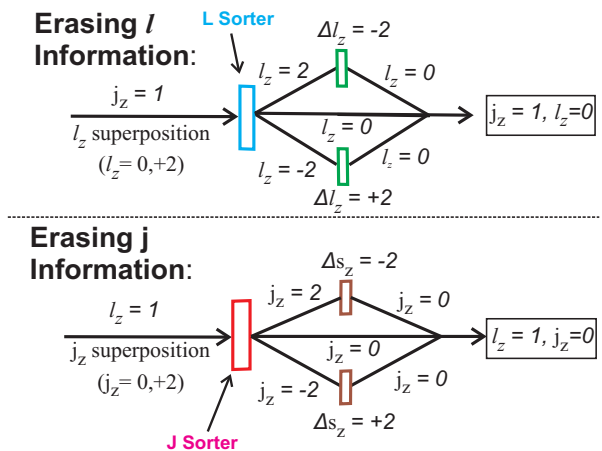}
\caption{\textit{The procedures for erasing $\bm {\hat L_z}$ information (top) following a $\bm {\hat J_z}$ measurement, or erasing
$\bm {\hat J_z}$ information (bottom) after an $\bm {\hat L_z}$ measurement. In the top case, the value of $\bm {\hat J_z}$ can still be determined
from which of an array of detectors
fires at the end (see figure \ref{detail}), and similarly for $\bm {\hat L_z}$ in the bottom case. The figure is drawn for incoming value of $l$ or $j$ equal to $1$, but the process works the same way for other values. In each case, the undesired values are sorted, shifted to zero, then recombined so
there is no way to determine what the original value was. The shifting is further illustrated in figure \ref{shiftfig}.
}}
\label{erasure}
\end{figure}

\begin{figure}
\centering
\includegraphics[totalheight=2.8in]{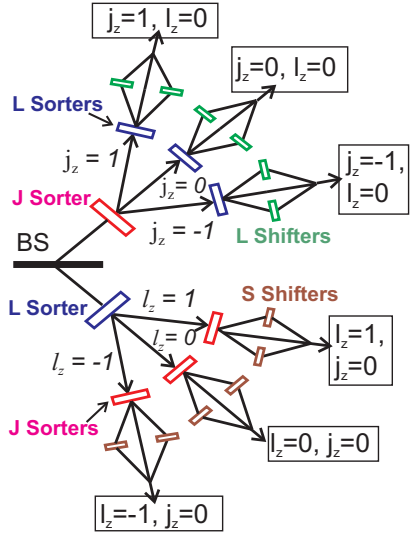}
\caption{\textit{A more detailed view of Alice's lab; Bob's lab has a similar arrangement.
After sorting in one variable (either $j_z$ or $l_z$), information about the other variable then needs to be erased (figure \ref{erasure})
before polarization interference is measured.
The outgoing arrows lead to systems for polarization measurement. For the top three
outputs in the figure, which particular detector fires at the end will tell Alice the value of $j_z$ but will give no information about $l_z$;
the opposite is true in the bottom three outputs. Thus Alice will know the value of only one of these two variables.}}
\label{detail}
\end{figure}

Figure \ref{sortfig} shows the proposed setup in schematic form, with more detailed view of portions of it in figures \ref{erasure} and
\ref{detail}. We will assume that the down conversion source uses a pump beam with zero OAM. The signal and idler polarizations are perfectly correlated for type
I down conversion or perfectly anticorrelated for type II; either way, the OAM is perfectly anticorrelated. For specificity, we henceforth assume type II down
conversion. The particular case drawn in figures \ref{erasure} and \ref{detail} assumes alphabet size $N=3$ ($l=0,\pm 1$; i.e. $l_0=1$), so an array of three
detectors is required following each sorter. Larger alphabets require more detectors, sorters, and erasure stages, but no further changes are needed; the setup
complexity therefore grows much more slowly with alphabet size than in other approaches; a change of OAM bases in the approach of \cite{grob}, for example,
requires the alignment and coordination of rapidly increasing numbers of moving stages as the dimension grows. (Note that even though the alphabet being used is
$\left\{ -1,0,+1\right\}$, the sorters will need to be capable of sorting values up to $\pm 2$ to carry out the erasure procedure in figure \ref{erasure}.)

\begin{figure}
\centering
\subfigure[]{
\includegraphics[width=.38\columnwidth]{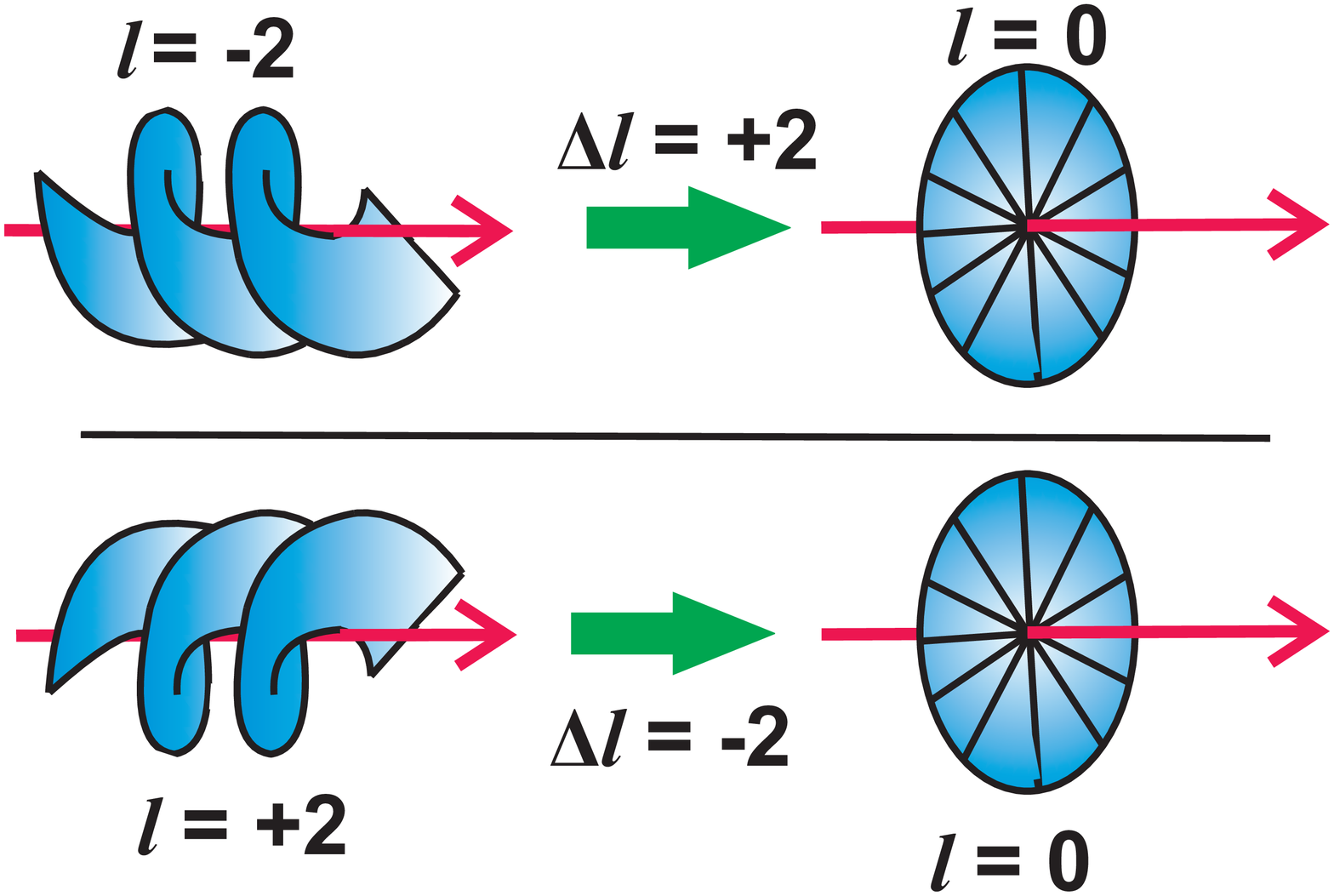}}
\subfigure[]{
\includegraphics[width=.34\columnwidth]{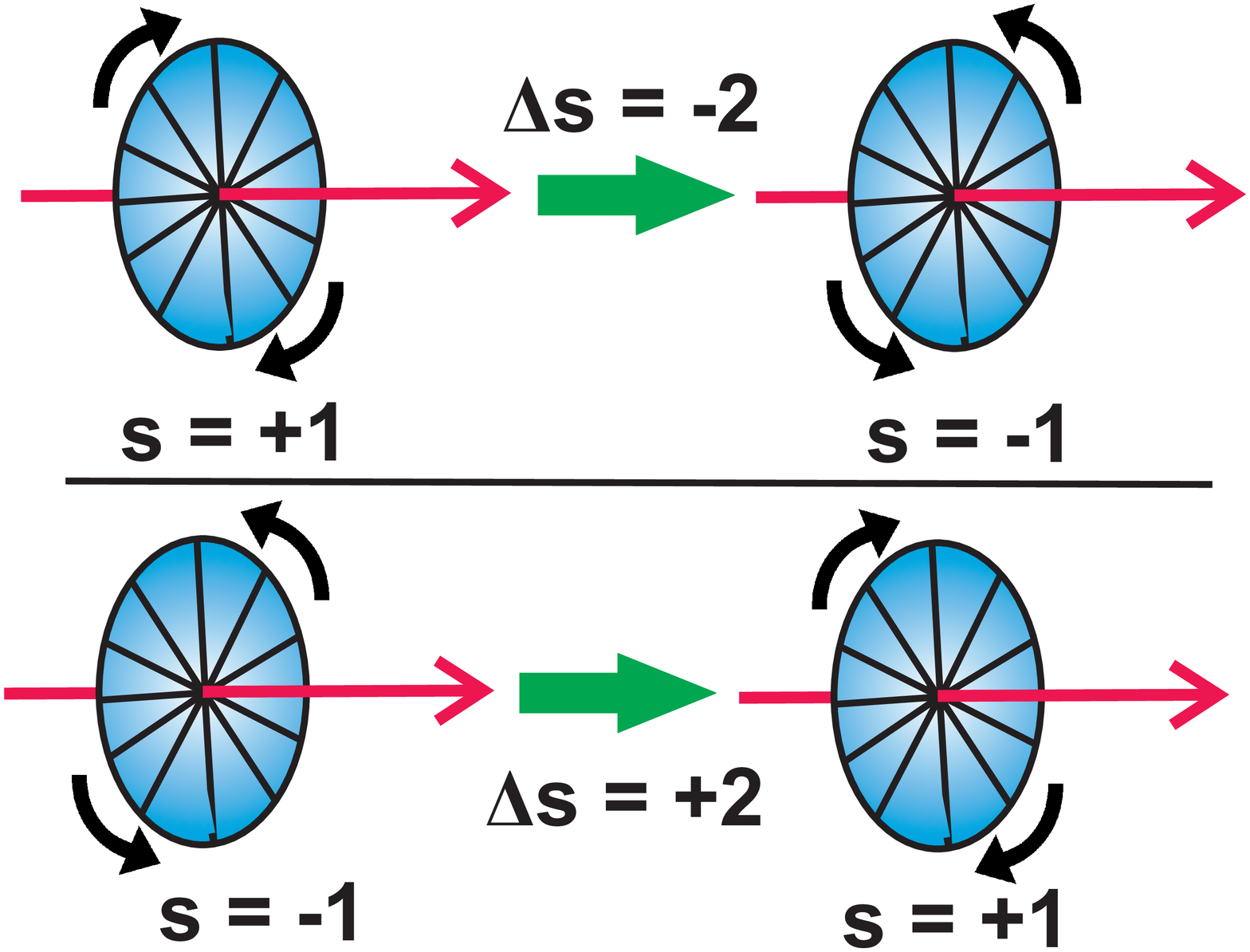}}
\caption{ The effect of the $l$ and $s$ shifters of figure \ref{erasure}.  (a) The shifting of $l$ in the top part of figure \ref{erasure} can be accomplished with a computer-generated hologram or
a spiral phase plate. The incoming waves of positive or negative OAM are converted into plane waves with $l_z=0$.
(b) The shift of $s$ in the bottom part of figure \ref{erasure} may be carried out with birefringent phase plates. Left- and right-circularly polarized waves are interchanged. Although this will alter the value of $s$, it will leave
the spin or polarization entanglement undisturbed on trials where both participants measure $\bm {\hat J_z}$, since they will \emph{both} carry out
similar spin flips in opposite directions.
}\label{shiftfig}
\end{figure}

Alice and Bob can readily measure either OAM eigenvalue $l_z$ or spin eigenvalue $s_z$ (circular polarization), or both. In the paraxial case, $\hat {\bm S}$ and
$\hat {\bm L}$ are well-defined and commute, so their components can be simultaneously measured, as verified experimentally in \cite{leach1}. This fully
determined $j_z$. In contrast, the TAM $j_z$ about the propagation axis can be measured interferometrically in such a way \cite{leach1} that it leaves the
separate values of \emph{both} spin and OAM undetermined. So suppose that Alice and Bob each have a beam splitter randomly sending incoming photons either to an
apparatus that measures $\hat {\bm L_z}$ or to one measuring $\hat {\bm J_z}$ (the sorters in figure \ref{sortfig}). After the sorting is done on one variable,
information about the other variable is erased by an arrangement of beam splitters, waveplates, and holograms (see figures  \ref{erasure}, \ref{shiftfig}, and
\ref{detail}). The sorting by $l_z$ and $j_z$ values is nondestructive, so the spin or polarization can still be measured afterwards.

\subsection{Erasing unmeasured variables} The erasure of the unmeasured variable, as illustrated in figure \ref{erasure}, is necessary because otherwise there will be no
interference between polarization states. To see this, imagine Alice and Bob place linear polarizers at respective angles $\theta_A$ and $\theta_B$ from the
horizontal. The states passed by the polarizers will be denoted $|\theta_A\rangle $ and $|\theta_B\rangle$. After passing through the polarizers, the probability
of joint detection in both labs is proportional to $|\langle \theta_A| \langle\theta_B|\psi \rangle |^2$. Provided the angles are not multiples of $\pi\over 2$,
both $|H\rangle$ and $|V\rangle$ will have nonzero projections onto the $|\theta\rangle $ states, so that cross-terms between the $H$ and $V$ pieces will survive
in the probability. These cross terms will be dependent on $\theta_A$ and $\theta_B$, giving rise to the desired Bell interference. However if the polarization
is entangled with another variable (OAM for example), a state such as $|l_1, H\rangle_A |l_2, V\rangle_B \pm |l_2, V\rangle_A |l_1, H\rangle_B$ will produce no
interference, since the $\bm {\hat L_z}$ eigenstates $|l_1\rangle $ and $|l_2\rangle$ will still be orthogonal after the polarizer, causing the cross terms to
vanish; there are no intermediate states to bridge the two orthogonal OAM states in the way that the $|\theta\rangle $ states did for polarization. This can be
seen in detail in the example given below (section \ref{example}).

When $j_z$ has been measured, information about $l_z$ can be erased (top part of figure \ref{erasure}) by sorting different $l_z$ values into different paths,
inserting appropriate holograms or spiral phase plates to shift the OAM in each path to zero, then recombining the paths. In this manner, the initial OAM values
are erased (shifted to zero) so that there is no way of determining which path was taken and what the initial OAM value was. The different incoming OAM states
are now indistinguishable, while the polarization states are left entangled. Similarly, when $l_z$ is measured information about $j_z$ can be erased (bottom part
of figure \ref{erasure}) by sorting $j_z$ values, shifting the values of $s_z$ appropriately with phase plates (converting one circular polarization into the
other), and recombining. Although this changes the value of $s$ for each photon, it leaves the entanglement undisturbed on the trials where both Alice and Bob
measure $j_z$, since they both carry out similar shifts: an incoming entangled spin state of the form $|s_z=1\rangle_A |s_z=-1\rangle_B \pm |s_z=-1\rangle_A
|s_z=1\rangle_B$ is shifted to $|s_z=-1\rangle_A |s_z= +1\rangle_B \pm |s_z=+1\rangle_A |s_z=-1\rangle_B$, which is still entangled and in fact proportional to
the original state.

\subsection{Procedure}

Consider now the setup described in the previous section with a two-photon input state generated from type II parametric down conversion. The pump beam is assumed to have no OAM, $l_{pump}=0$.   Alice and Bob each receive one photon from the state, on which to make measurements. Consider several possibilities:

$\bullet$ Suppose Alice and Bob both measure $l_z$. Using the fact that their values should be perfectly anticorrelated, they can use the resulting OAM quantum
number on Alice's side (or, equally, on Bob's) to define a key. Since $\hat {\bm L}$ and $\hat {\bm S}$ measurements don't affect each other, the spin
eigenvalues along each axis remain undetermined and the polarization state remains entangled.

$\bullet$ Alternatively, if both measure $j_z$, the key can then be defined by the resulting TAM quantum number on Alice's side. The spin components are again
undetermined, and the polarization state remains entangled.

$\bullet$ But if one measures $j_z$ and the other measures $l_z$, this completely fixes the spin along the axis: $s_z=j_z-l_z$. The spin wavefunction collapses
from an entangled state to a separable one. These trials are discarded.

In the first two cases, the spin states remain entangled after sorting, so tests on the linear polarization should yield Bell violation. In the third case, when
Alice and Bob measure \emph{different} variables, such a test should yield no violation, the spin having been reduced to a classical quantity. The measurement of
one variable ($\hat {\bm J_z}$ or $\hat {\bm L_z}$) reduces the original space of states for each particle to a two-dimensional subspace, while measurement of
the second variable further reduces each particle to one unique state. Consequently, the two-photon pair goes from an entangled to a separable state.

If Eve attempts quantum nondemolition measurements to determine $j_z$ or $l_z$, then on half the retained trials she will measure the wrong variable (the one not
measured by Alice and Bob), thus fixing $s$ and causing a detectable loss of Bell violation.

In this scheme, the beam splitter's random choice between causing either a $\hat {\bm J_z}$ measurement or an $\hat {\bm L_z}$ measurement replaces the usual
random modulation between two measurement bases for components of a single variable. All variables act in a completely quantum manner, with the "quantumness"
comes from the fact that although $\hat {\bm L}$ and $\hat {\bm S}$ commute with each other, neither commutes with $\hat {\bm J}$; if $j_z$ and $l_z$ are both
measured, whether by the legitimate participants or by eavesdroppers, polarization entanglement is destroyed.

Another equivalent way to view the situation, that makes the analogy to the Ekert case clearer, is that measurements can be made along a basis in angular
momentum space aligned with the $\hat{\bm J_z}$ axis or one aligned with the $\hat{\bm L_z}$ axis; these are mutually unbiased on each two-dimensional subspace
defined by a fixed $j_z$ value or a fixed $l_z$ value, but are also incomplete in the sense that neither measurement fully determines the state. However making
\emph{both} measurements \emph{does} uniquely determine the state, completely fixing $j_z$, $l_z$, and $s_z$ values.

Note that the key-generating capacity grows with increasing dimension, as is the case for all OAM-based QKD methods. However, unlike in other OAM-based schemes,
the security-checking remains essentially two-dimensional so that the level of security grows more slowly with increasing dimension. This is the price that is paid for
reducing the complexity of generating practical high-dimensional keys. The secure key rate and mutual information between participants will be examined in
section \ref{securitysection} and the appendix.

\subsection{Example}\label{example}

To be more concrete, consider an entangled two-photon input state of the form
\begin{eqnarray}|\psi\rangle &=&  |\psi_{oam}\rangle |\psi_{spin}\rangle \\ &=& {1\over
\sqrt{2}}\sum_{l_z=-l_0}^{l_0} |l_z\rangle_A|-l_z\rangle_B \Bigl( |H\rangle_A|V\rangle_B -|V\rangle_A |H\rangle_B\Bigr) .\label{state1}\end{eqnarray} Such a
state arises, for example, from type II down conversion after filtering to equalize the probabilities of various $l$ values. The linear and circular polarization
states are related by \begin{eqnarray}|H\rangle &=& {1\over \sqrt{2}}\left( |s_z=1\rangle + |s_z=-1\rangle \right) \; =\; {1\over \sqrt{2}}\left( |R\rangle
+|L\rangle \right) \\  |V\rangle &=& {i\over \sqrt{2}}\left( |s_z=1\rangle - |s_z=-1\rangle \right) \; =\; {i\over \sqrt{2}}\left( |R\rangle -|L\rangle \right)
,\end{eqnarray} where $R$ and $L$ correspond to spin $s_z=+1$ and $s_z=-1$, respectively. So the two-particle state $|\psi\rangle$ can be written in terms of
joint OAM-spin states $|l,s\rangle $ as
\begin{eqnarray}|\psi\rangle = -{1\over\sqrt{2}}\sum_{s_z=-1}^1 \sum_{l_z=-l_0}^{l_0} (-1)^{s_z/2} |l_z,s_z\rangle_A |-l_z,-s_z\rangle_B  . \end{eqnarray} If both
experimenters measure the TAM with Alice obtaining value $j_z$, then Bob will obtain value $-j_z$,
reducing the state to \begin{eqnarray}|\psi^\prime \rangle &=& {-i\over \sqrt{2}} \left( |j_z-1,+1\rangle_A |-j_z+1,-1\rangle_B \right. \nonumber\\
& & \qquad \left.-|j_z+1,-1\rangle_A |-j_z-1,+1\rangle_B\right)  .\label{psi_j_measured}\end{eqnarray} The first entry in each ket is the $l_z$ value, the second
is $s_z$. Note that $l_z$ and $s_z$ remain indeterminate after the $j_z$ measurement since multiple combinations of $l_z$ and $s_z$ can add up to the same $j_z$.
If $l_z$ is not measured in either branch at any point, then information about it can be erased as in figure \ref{erasure}, in order to arrive at a
maximally-entangled spin wavefunction,
\begin{eqnarray}
|\psi_{spin}\rangle &=&  -{i\over \sqrt{2}} \left( |R\rangle_A |L\rangle_B - |L\rangle_A |R\rangle_B  \right)\\
&=& {1\over \sqrt{2}}\left( |H\rangle_A |V\rangle_B - |V\rangle_A |H\rangle_B  \right) \label{psispin} .\end{eqnarray} Carrying out a Bell-type test on
polarization after the $j_z$-sorting then yields maximal quantum-mechanical Bell violation. However, if in addition to the $j_z$-sorting, the OAM on Bob's side
is also measured (by Bob or by Eve), the state of equation (\ref{psi_j_measured}) collapses  to either
\begin{equation}i|j_z-1,1 \rangle_A |-j_z+1,-1\rangle_B \end{equation} (if Bob finds value $l_z=-j_z+1$), or else to \begin{equation} -i
|j_z+1,-1\rangle_A |-j_z-1,1\rangle_B\end{equation} (if Bob's value is $l_z=-j_z-1$). Placing quarter-wave plates at the output, to convert from circular to
linear polarization, the state becomes either\begin{eqnarray} i |j_z-1,H\rangle_A |-j_z+1,V\rangle_B\end{eqnarray} or \begin{eqnarray} -i |j_z+1,V\rangle_A
|-j_z-1,H\rangle_B .\end{eqnarray} Either way, it is now a separable state (both before and after the $l_z$ erasure) with definite polarization for each photon,
so no Bell violation occurs. If Alice and Bob measure $l_z$ while Eve measures $j_z$, a similar result follows.

When Eve guesses whether to measure $\hat {\bm L_z}$ or $\hat {\bm J_z}$, half the time she guesses wrong and causes collapse of the entangled polarization state
into a separable state. This lowers the interference pattern's visibility to classical levels when Bell tests are performed, providing a clear signal of her
intervention.

To examine the interference visibility, define the rotated polarization states at Alice's location:
\begin{eqnarray}|\theta \rangle_A &=& \cos \theta |H\rangle_A +\sin \theta |V\rangle_A \\
|\theta^\perp \rangle_A &=& -\sin \theta |H\rangle_A +\cos\theta |V\rangle_A ,\end{eqnarray} with similar states $|\phi\rangle_B$ and $|\phi^\perp\rangle_B$
defined at Bob's lab. $\theta$ and $\phi$ are the angles of linear polarizers before Alice's and Bob's detectors, respectively. In the absence of eavesdropping,
it is straightforward to verify that under ideal conditions (perfect detectors and no losses) the coincidence rate is proportional to \begin{equation}|\langle \psi_{spin}|\theta \rangle_A |\phi \rangle_B |^2\; =\; {1\over 2}\sin^2(\theta-\phi )\; =\; {1\over
2}\left[ 1-\cos^2(\theta -\phi )\right]\end{equation} for the two-photon entangled spin state of equation (\ref{psispin}). A Clauser-Horne-Shimony-Holt (CHSH)-type interference experiment \cite{chsh}  will then exhibit oscillations with visibility ${\cal V}$ of $100\%$. On the other hand, if $|\psi_{spin}\rangle$ is replaced by any separable state of the form $|\psi_{sep}\rangle=|\gamma\rangle_A|\gamma^\perp\rangle_B$ (where $\gamma$
is the polarization direction of the photon measured by Alice), the corresponding inner product is $|\langle \psi_{sep}|\theta \rangle_A |\phi \rangle_B |^2 =\cos^2 (\gamma-\theta )\sin^2(\gamma -\phi ); $ the dependences on $\theta $ and $\phi$ now factor so that the visibility can never be greater than the classical limit of ${1\over \sqrt{2}}\approx 71\%$.  In general, if Eve is eavesdropping a
fraction $\eta$ of the time, the visibility will be ${\cal V}\le 1-\left( 1-{1\over \sqrt{2}}\right) \eta $. A drop in visibility to below the value set by the Bell-CHSH inequality signals the possible presence of eavesdropping.

\section{Information and security considerations}\label{securitysection}

Instead of testing Bell inequalities on the entangled polarization states, there is a second way to check the security of the transmission, which will be more
useful for arriving at quantitative estimates of information and signalling rates. On the set of trials for which Alice and Bob measure the same variable, they
can choose a random subset of their measurement values ($l_z$ or $j_z$ values) for comparison. Ideally, they should both find perfectly anticorrelated values, so
that the presence of discrepancies beyond the expected error rate due to the transmission method then serves as a signal of an eavesdropper's presence. This
method is more directly analogous to that of the BB84 scheme, requiring {\it no} active modulation of the detector settings, as opposed to the
Bell-inequality-based version of the Ekert scheme, which still requires modulation of the settings for the polarization measurements. In this section, we take
the BB84-like approach when we look at security considerations, since it is relatively easy to compute the probabilities of the output states and the key rate
while taking Eve's actions into account.

If the allowed OAM values are $\left\{-l_0,\dots ,0,\dots ,+l_0\right\}$, then the possible $j_z$ values are $ \left\{-l_0-1,\dots ,0,\dots ,l_0+1\right\} $. It
should be noted that unlike the case of the Ekert or BB84 protocols, where there is a finite number of possible outcomes (two polarizations) and they are both
used, in the current case we make use of a finite subset of a larger (in fact infinite) set of possible output values for $j_z$ or $l_z$. As a result, when Eve
interferes it is possible to "run off the end" of the allowed set of values, and this possibility must be accounted for. Also, we note that there are $2l_0+3$
values of $j_z$ but only $2l_0+1$ values of $l_z$. As a result of these complications some further refinements to the method must be made. These are discussed in
the appendix. The appendix also then gives the resulting probability distributions for Alice's and Bob's joint outcomes. In this section, we make use of those
distributions to investigate the quantum security of the procedure against eavesdropping.

\begin{figure}
\centering
\includegraphics[totalheight=2.4in]{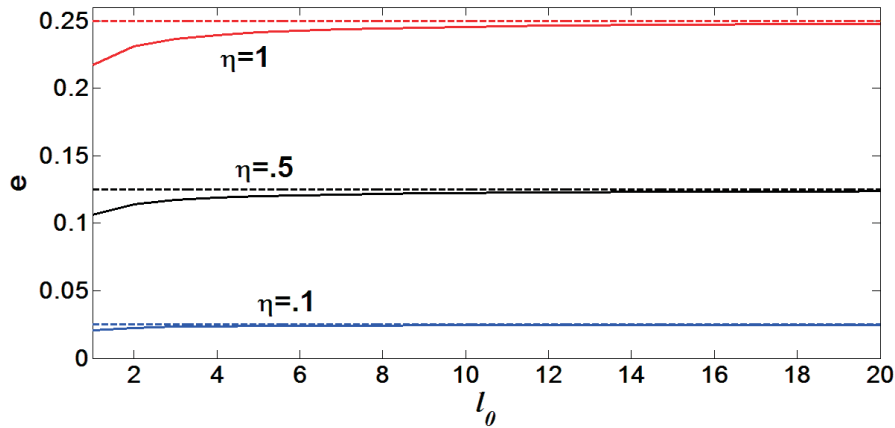}
\caption{\textit{Comparison of the error rate $e$ for the scheme described in this paper (solid curves) to the BB84 protocol (dashed lines).
This is done for (from top to bottom) eavesdropping ratios $\eta=1$, $\eta=.5$, and $\eta=.1$. In each case, the error rate for the current approach
starts a little below the BB84 value for small $l_0$, but approaches it rapidly  and asymptotically as $l_0$ increases.}}
\label{errorfig}
\end{figure}

\subsection{Error rates}
If Eve intercepts fraction $\eta$ of the transmissions, she has  $50\% $ chance of measuring the correct variable each time, obtaining the correct key value $k$
(either from an $\hat {\bm L_z}$  or $\hat {\bm J_z}$ measurement) without introducing errors. On the other hand, during the $50\%$ of the times in which she
measures the wrong variable, she only has a further $50\% $ chance of sending the correct value to Bob. Specifically, each time she measures the wrong variable
her intervention has probability ${1\over 8}$ of causing Bob to measure the value $k-2$ and overall probability ${1\over 8}$ of causing him to measure the value
$k+2$, with only probability ${1\over 2}$ of obtaining the correct key value, $k$ (see figure \ref{splittingfig}). So if the alphabet size were infinite, the
eavesdropper-induced error rate would be $e=\eta \cdot\left( {1\over 2} \right)\cdot \left({1\over 2} \right) \; =\; {\eta\over 4}.$ However, we must take into
account the fact that Eve's actions can cause values to move out of the range being used for the key; this will alter the error rate slightly. Using the
probability distribution $P_{AB}$ given in the appendix, it is then straightforward to show that the true eavesdropper-induced error rate is
\begin{equation}e={\eta\over 8}\left( {{4l_0+1}\over {(2l_0+1)-{\eta\over 8}}}\right).\end{equation} Bob's error rate is shown in figure \ref{errorfig} for three
different eavesdropping ratios, along with the corresponding values for polarization-based BB84. It is seen that for small $l_0$ the error rate is slightly lower
than the BB84 value, but it rapidly approaches that value as $l_0$ increases.

After dropping the trials on which Alice and Bob measure different variables, the fraction $f$ of the remaining photons that are used to generate the key may also be easily computed from $P_{AB}$. It is found to be
\begin{equation}f={{4l_0+2-\eta}\over {4l_0+3}}.\end{equation} This approaches $100\%$ for $l_0\to \infty$. For finite $l_0$, it ranges
between a low of $f={{4l_0+1}\over {4l_0+3}}$ (for $\eta =1$) and a high of $f={{4l_0+2}\over {4l_0+3}}$ (for $\eta =0$). For the worst
case ($l_0=1$ qutrits), this corresponds to a range of ${5\over 7}$ to ${6\over 7}$.

\subsection{Mutual information and key rates}

\begin{figure}
\centering
\subfigure[]{
\includegraphics[width=.48\columnwidth]{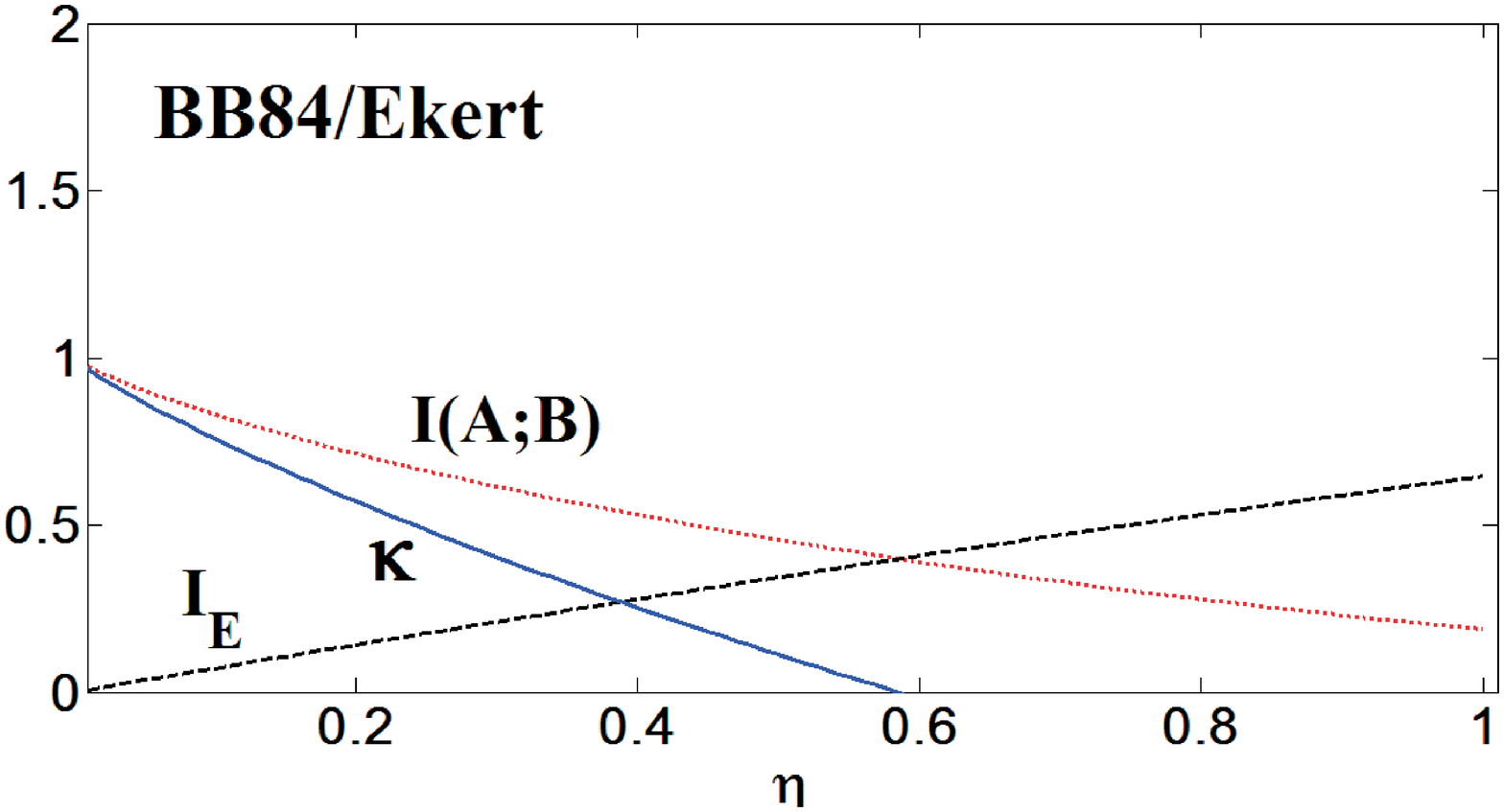}}
\subfigure[]{
\includegraphics[width=.48\columnwidth]{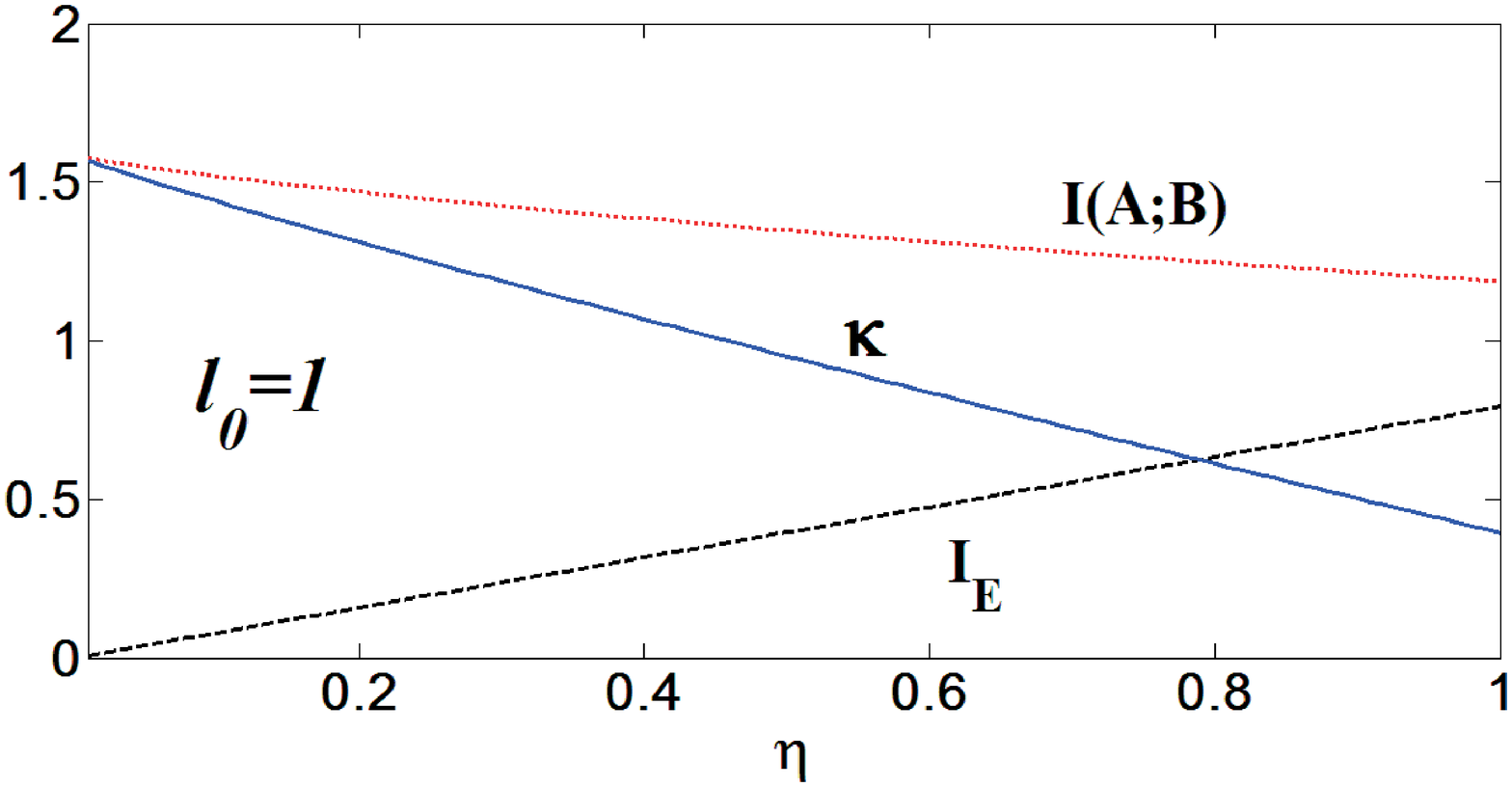}}\\
\subfigure[]{
\includegraphics[width=.48\columnwidth]{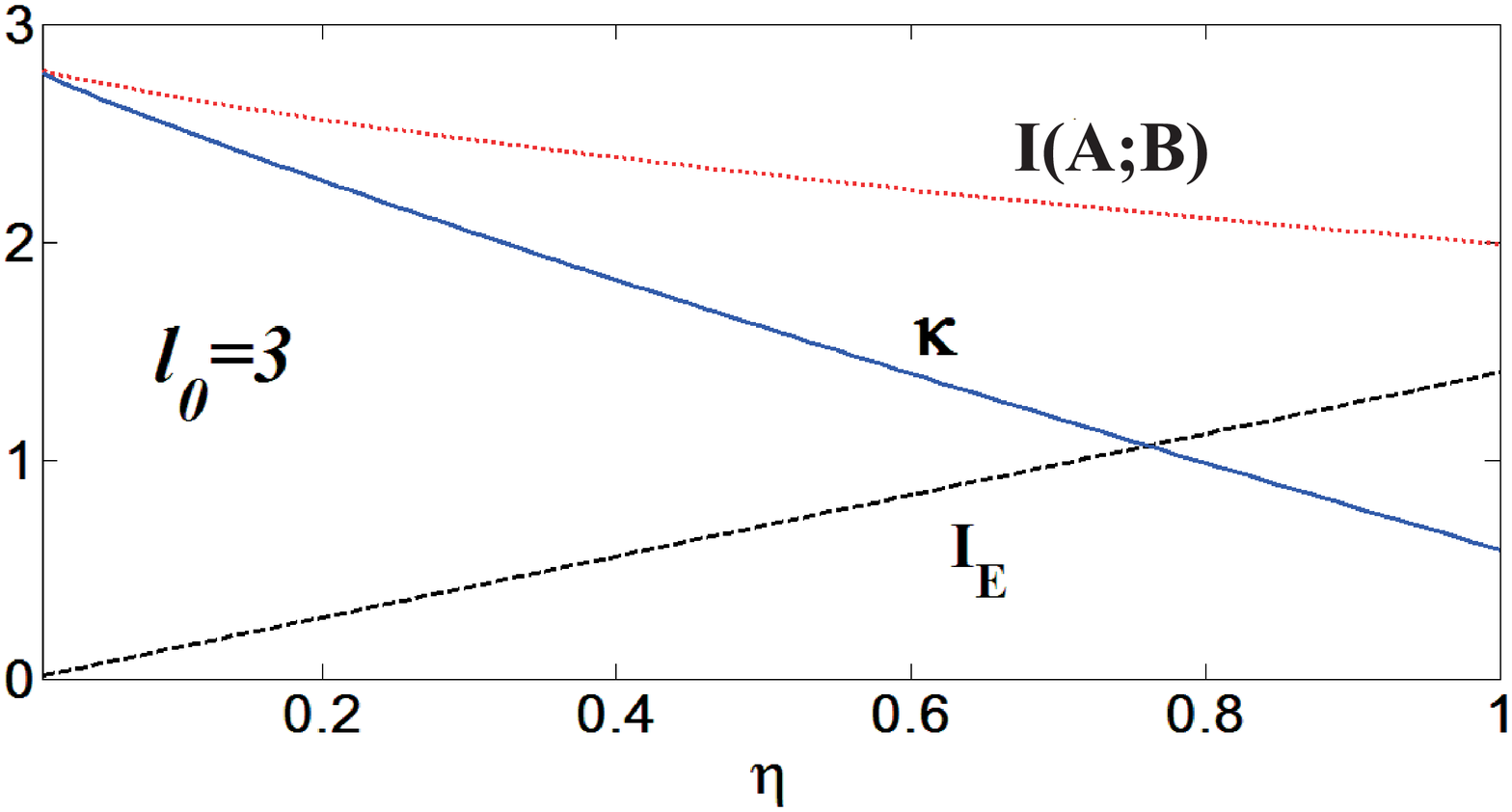}}
\subfigure[]{
\includegraphics[width=.48\columnwidth]{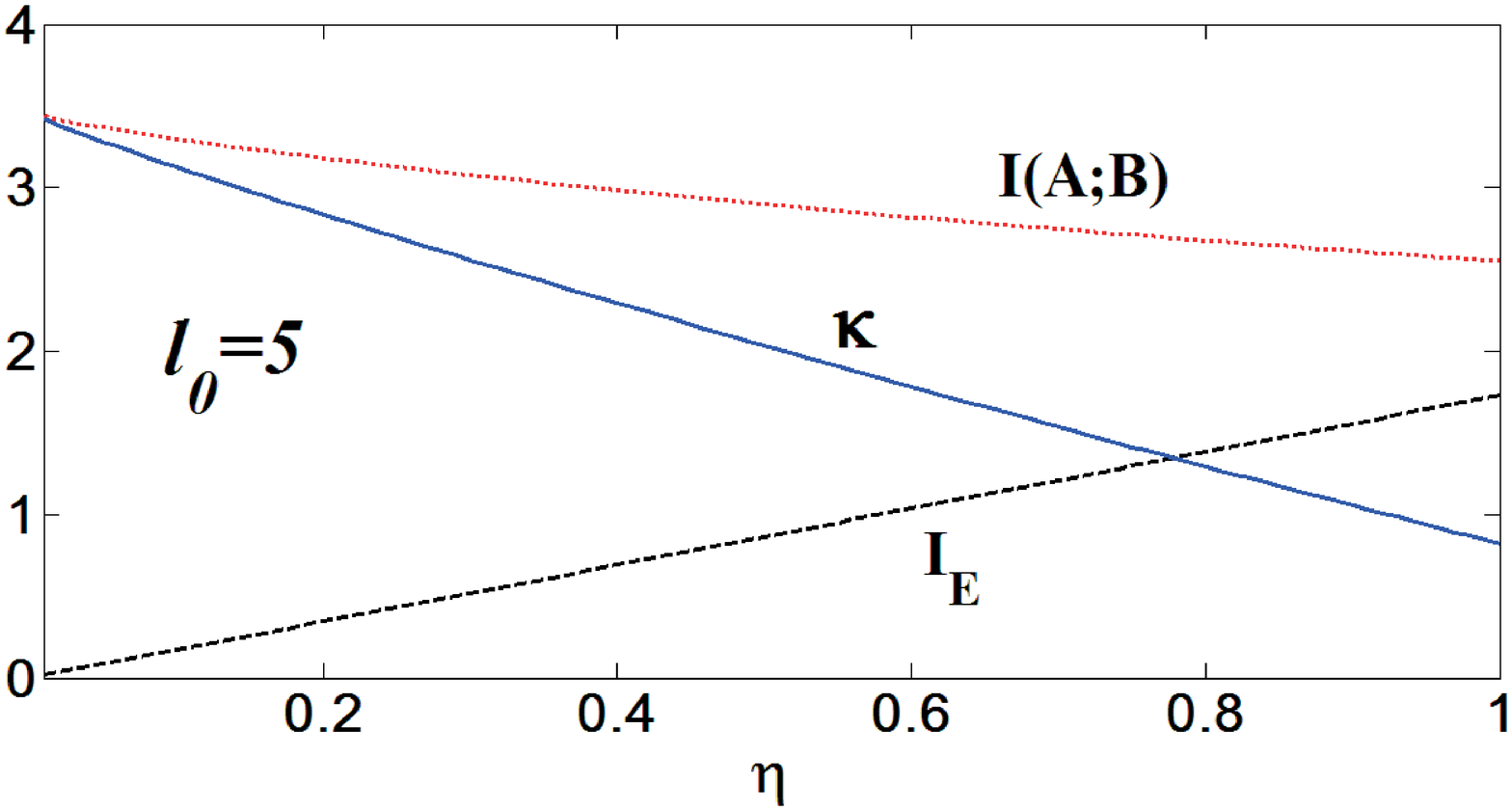}}
\caption{Mutual information between Alice and Bob $I(A;B)$ (dotted red curve)
and information gain by Eve $I_E$ (dashed black) for (b) $l_0=1$, (c) $l_0=3$, and (d) $l_0=5$.
The secret key rate $\kappa$ (solid blue) is either the difference $I_{AB} - I_{AE}$, or zero,
whichever is larger. As long as $\kappa >0$, a secret key can always be distilled. For comparison, the same
quantities are plotted for the polarization-based Ekert protocol (or equivalently, for the BB84 protocol) in (a).
(Note the change in scale on the vertical axes of the different plots.)}\label{infofig}
\end{figure}

One may also compute the mutual information between the legitimate agents, $I(A;B)$, and information gain of the eavesdropper, $I_{E}=\mbox{max}\left\{ I(A;E),I(B;E)\right\}$. From these, the secret key rate, $\kappa =max\left\{ I_{AB}-I_E,0\right\}$ may be found.
Recall that it is always possible to distill a secret key using privacy amplification when $\kappa >0$. Instead of the distribution $P_{AB}$ given in the appendix (which includes \emph{all} events, even those for which the values run off the edge of the alphabet and so generate no key) in order to compute $\kappa$, we must use the probability distribution $P_K(A,B)$ for the \emph{key-generating events only}. This new distribution is obtained from  $P_{AB}$ simply by dropping its last row and column, then dividing by the key-generating fraction $f$ in order to renormalize the total probability to unity. Straightforward calculation then gives the result that the mutual information \cite{cover} between Alice and Bob as a function of parameters $\eta$ and $l_0$ is:
\begin{eqnarray}I(A;B)&=& {2\over {4l_0+2-\eta}}\left\{ \left( 2l_0+1-{\eta\over 2}\right)\log_2\left( {{4l_0+2-\eta}\over 2}\right)
\right. \\
& & \quad
-8\left( 1-{\eta\over 8}\right) \log_2\left( 1-{\eta\over 8}\right)  + (2l_0+1)\left( 1-{\eta\over 4}\right)\log_2\left( 1-{\eta\over 4}\right) \nonumber \\
& & \left. \quad +{\eta\over 4}(2l_0-1)\log_2 {\eta\over 8}\right\} \nonumber \end{eqnarray} Asymptotically (for $l_0\to \infty$) this approaches $\log_2(2l_0)$,
independent of $\eta$; for finite $l_0$ and no eavesdropping ($\eta =0$), it is equal to $\log_2(2l_0+1)$. Eve gains full information about the key value on half
of the measurements she makes and receives none on the other half, so the information gained by Eve is simply ${\eta\over 2}$ times the information per photon.
The results for the information and the secret key rate are plotted in figure \ref{infofig}(b-d). The case of BB84 is shown for comparison in figure
\ref{infofig} (a). It is seen that $\kappa$ is always greater than in the BB84 case, that it remains positive for all values of $\eta$, and that for any fixed value of
$\eta$ the value of  $\kappa$ increases with increasing $l_0$. Thus, the amount of key generated per transmitted photon is significantly larger than in the
BB84 or Ekert schemes. Security can be further enhanced in various ways, such as replacing the four-state polarization scheme with a six-state approach
\cite{bruss}.

%

\section{Conclusion} We have proposed a method for performing QKD using high-dimensional OAM and TAM variables in a manner that does not require the
complicated high-dimensional basis modulations necessary in other approaches, and which allows an increase in the rate of secure key distribution. The main
ingredients are: (i) A hyperentangled system with different functions segregated into different entangled degrees of freedom.  (ii) Random switching between two OAM \emph{bases} is replaced by random switching between measurements of two distinct noncommuting but related \emph{variables}, $\hat {\bm L_z}$ and $\hat {\bm J_z}$. (iii) Measurement
of any one variable does not completely determine the state, while measurement of any two of the three relevant variables does. Together, these ingredients allow
the high capacity of $\bm{\hat J_z}$ or $\bm{\hat L_z}$ eigenstates to be used while modulating only the simpler polarization states. As a result,
higher-dimensional OAM spaces can be utilized and higher key-generation capacities can be achieved with only relatively minimal increases in apparatus
complexity.

Previous approaches to using polarization and OAM together simply used them to generate larger keys from each photon by increasing the number of variables
involved. In these past approaches, the variables still remained separate, with no interplay between them. QND measurements on one variable left no signature in the other, so that both must experience independent basis modulations to maintain security. In contrast, the approach described here makes a more
fundamental use of the system's hyperentanglement, constructing a chain of three variables such that adjacent pairs in the chain do not commute. This noncommutativity
provides a linkage between the variables that enhances security, as well as increasing the number of key bits per photon.

The approach of constructing chains of
pairwise noncommuting operators has not been previously used in QKD and is likely to be generalizable to other operators (aside from angular momentum), and to employment in other types of quantum
protocols as well.
The procedure also makes use of an angular momentum erasure process that is of interest in its own right, since it allows the possibility of conducting future
quantum erasure or delayed choice experiments  in angular momentum space.

At high $l_0$, the eavesdropper-induced error rate is lower than for other two-basis OAM-based QKD schemes, where $e={\eta\over 2}\left( 1-{1\over
{2l_0+1}}\right)=\eta {{l_0}\over {2l_0+1}}$; in the current scheme $e$ instead remains near the 2-dimensional BB84 error rate; this is because the eigenspaces
of the two measured operators ($\bm {\hat J_z}$ and $\bm {\hat L_z}$) are effectively unbiased only on a two-dimensional subspace, due to the two-dimensional
nature of the polarization. The dimension of the subspace on which the variables are unbiased does not increase with $l_0$. Because of this, the secret key rate
$\kappa$ will be lower than for other OAM-based schemes at large $l_0$. However, $\kappa$ in this approach is always higher than in BB84 or E91 protocols even
for the least advantageous case ($l_0=1$), and it grows logarithmically with increasing $l_0$; similarly, the BB84-level eavesdropper-induced error rate remains
sufficient to detect eavesdropping regardless of dimension.

There are some technical difficulties that must be overcome for the method to become practical. Probably the chief among these is that the most common method of measuring photon OAM is to shift the input $l$-state to $l = 0$ and then to collect them in an optical fiber. This method is of low efficiency, which greatly reduces the key transmission rate of all OAM-based schemes. Further, the interferometers used to sort the OAM and TAM values \cite{leach1} become progressively more complex as the range of $l_z$ and $j_z$ values to be used increases.

However, when coupled to the much greater ease in this scheme of switching between $\hat J_z$ and $\hat L_z$ measurements compared to the difficult switching between measurements of different components of $\hat{\bm L}$ in other protocols, the method
presented here seems to hold strong promise as a more practical way to reach higher key rates per photon while maintaining full quantum-level security.

\ack This research was supported by the DARPA QUINESS program through US Army Research Office award W31P4Q-12-1-0015, by the DARPA InPho program through US Army
Research Office award W911NF-10-1-0404, and by National Science Foundation grant ECCS-1309209.  The authors would like to thank Prof. Daniel Gauthier, Dr. Thomas
Brougham, and Dr. Kevin McCusker for helpful comments concerning an early version of this manuscript.

\appendix

\section{Outcome Probabilities}
As mentioned in section \ref{securitysection}, we must make adjustments to the protocol in order to equalize the probabilities of the allowed key values,
and must
take into account that the values measured may lie outside the range used for key generation. We deal with those complications here, and give the joint
probability distributions that result after doing so.

\subsection{Undisturbed probability distributions}
First, the key must have the same range of values regardless of which variable was measured. So assign no key value when $j_z=\pm (l_0+1)$ is measured; for the
remaining $j_z$ values, assign key value $j_z$. Then either type measurement leads to a range of values from $-l_0$ to $l_0$, leading to alphabet size
$N=2l_0+1$. Instances where $j_z$ equals $l_0-1$ or $l_0+1$, though not used for key generation, are not discarded; they are recorded for use in the security
analysis. Second, all key values must have equal probability. Initially, each $l_z$ value has probability $P_L(l_z)={1\over {2(2l_0+1)}}$ (the $1\over 2$ comes
from the probability that $l_z$ was measured instead of $j_z$), while each $j_z$ used for key generation has probability $P_J(j_z)={1\over {2(2l_0+2)}}$. (The
two values not used for key generation each have probability ${1\over {4(2l_0+3)}}$.) To make $P_L(k)=P_J(k)$ for each key value $k$, the reflectance of the beam
splitter may be adjusted away from $50\% $, so the reflection and transmission probabilities are $|r|^2={1\over 2}-\epsilon$ and $|t|^2={1\over 2}+\epsilon$,
with $\epsilon= {1\over 2}\left( {1\over {4l_0+3}}\right).$ Then each $l_z$ and $j_z$ value has probability
\begin{equation}P_L(l_z)=P_J(j_z)={1\over {4l_0+3}},\end{equation} and each possible key value has probability \begin{equation}P(k)=P_L(k)+P_J(k)=
{2\over {4l_0+3}},\end{equation} with probability ${1\over {4l_0+3}}$ that no key is generated.

\begin{figure}
\centering
\includegraphics[totalheight=2.4in]{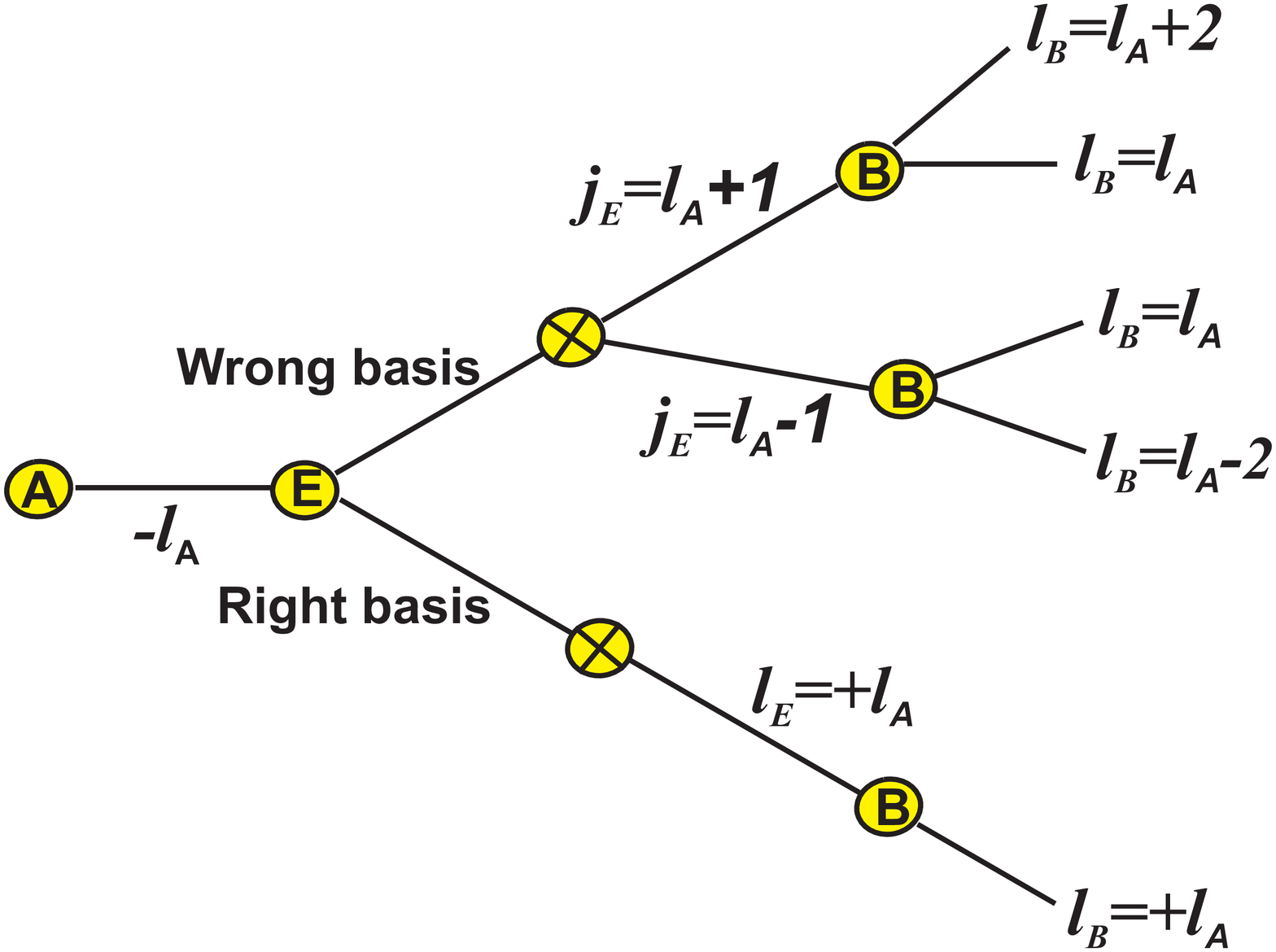}
\caption{\textit{The possible outcomes when Eve intervenes, assuming Alice and Bob both measure $\hat
{\bm L_z}$. (If they measure $\hat {\bm J_z}$ instead, swap the variables $l$ and $j$ everywhere in the figure.)
At each splitting, each branch has equal probability. Circles containing the letters A, B, or E represent measurements by Alice, Bob, or Eve, respectively.}}
\label{splittingfig}
\end{figure}

In Eve's absence, there is ideally perfect anticorrelation when Alice and Bob
measure the same variable. The distributions of key values $k$ should be identical,
$P_A(k)=P_B(k).$ Moreover, their joint distribution should be
uniform on the diagonal and vanishing elsewhere: \begin{equation}P_0(k_A,k_B)=\left( {2\over {4l_0+3}}\right)
\delta_{k_Ak_B}.\end{equation} We therefore find entropies $H(A) = H(B) = H(A,B) =
 \log_2(4l_0+3) -\left( 4l_0+2\right)/\left( 4l_0+3\right) ,$ so the mutual
information $I(A;B)\equiv H(A)+H(B)-H(A,B)$ just equals the Shannon information of each participant separately,
$I=\log_2(4l_0+3) -\left( 4l_0+2\right) / \left(4l_0+3\right) .$

\subsection{Probability distributions with eavesdropping}
The effects of Eve's actions are shown in figure \ref{splittingfig}. (The figure assumes that Alice and Bob measure $\bm {\hat L_z}$; if they measure $\bm {\hat
J_z}$, simply interchange $l_z$ and $j_z$ everywhere in the figure.) At each splitting of branches, there is a $50\%$ chance that each branch will be taken.
Suppose Alice measures $\bm {\hat L_z}$ and obtains value $-l_A$. If Eve intercepts the transmission, she may measure the same variable, in which case both she
and Bob will obtain the negative of Alice's value: $l_E=+l_A$ for Eve and $l_B=+l_A$ for Bob. But if Eve measures the other variable, $\bm {\hat J_z}$, then she
has equal likelihood of measuring the value above $l_A$ or the value below it: $j_E=l_A+1$ or $j_E=l_A-1$. Similarly, Bob then has equal chances of measuring
$\bm {\hat L_z}$ to have the eigenvalue one unit above or below Eve's value: $l_B=j_E\pm 1$.

As a result, the entries of the matrix representing the undisturbed joint probability distribution,
\begin{eqnarray}P_0(k_A,k_B) = \left( {2\over {4l_0+3}}\right) \delta_{k_Ak_B} \end{eqnarray} are now smeared
out by Eve's actions over multiple entries in Bob's direction.


Using the diagram of figure \ref{splittingfig}, the new Alice-Bob joint probability distribution on trials where Eve intervenes
may be determined. It is found to be
\begin{equation}P_1={2\over {4l_0+3}}\left( \begin{array}{cccccc|c}
{3\over 4} & 0& {1\over 8} & 0 & \dots & 0& {1\over 8}   \\
0 & {3\over 4} &  0 & {1\over 8} &  & &{1\over 8}\\
{1\over 8} & 0 & {3\over 4} &  0 & {1\over 8} &  &0\\
 & \ddots & \ddots & \ddots & & & \vdots \\
& & \ddots & \ddots & \ddots & &  0\\
& &  &  {1\over 8} & 0 &  {3\over 4} &   {1\over 8} \\ \hline
&\dots & 0& 0& {1\over {16}}& 0&  {7\over {16}}
\end{array}\right) ,\end{equation} where
rows label Alice's outcomes and columns label Bob's. The first $2l_0+1$ rows and columns label
possible key values, while the last row and column correspond to Alice or Bob, respectively,
generating no key.
For eavesdropping fraction $\eta$, the full joint
outcome distribution for all trials becomes \begin{equation}P_{AB}(k)=\left( 1-\eta\right) P_0+ \eta P_1 ,\end{equation} with
marginal probabilities for the two participants obtained by summing rows and columns.

The eavesdropper-induced error rate, the mutual information shared by Alice and Bob, and the secure key rate
may all now be found using this distribution. These quantities are discussed in section \ref{securitysection}.

\end{document}